\documentclass[useAMS,usenatbib]{mn2e}
\usepackage{amssymb}		
\usepackage{epsfig}
\def\bk{{\bf k}}
\def\bx{{\bf x}}
\def\bp{{\bf{p}}}
\def\bq{{\bf{q}}}
\def\beqra{\begin{eqnarray}}
\def\eeqra{\end{eqnarray}}
\def\beq{\begin{equation}}
\def\eeq{\end{equation}}
\def\vp{\varphi}

\title[Halo clustering with resummed perturbation theories]{Modeling the
clustering of dark-matter haloes in resummed perturbation theories}
\author[A. Elia, S. Kulkarni, C. Porciani, M. Pietroni, S. Matarrese]{A. Elia$^{1}$\thanks{E-mail:
elia@astro.uni-bonn.de}, S.
Kulkarni$^{2}$, C. Porciani$^{1}$, M. Pietroni$^{3}$, S. Matarrese$^{4,3}$ \\
$^{1}$Argelander Institut f\"ur Astronomie der Universit\"at Bonn, Auf dem
H\"ugel 71, D-53121 Bonn, Germany\\
$^{2}$Bethe Center for Theoretical Physics \& Physikalisches Institut,
Universit\"at Bonn, Nu\ss allee 12, D-53115
Bonn, Germany\\
$^{3}$INFN, Sezione di Padova, Via Marzolo 8, I-35131 Padova, Italy\\
$^{4}$Dipartimento di Fisica ``G. Galilei'', Universit\`a degli Studi di Padova, Via Marzolo 8, I-35131 Padova, Italy}
\begin{document}
\pagerange{\pageref{firstpage}--\pageref{lastpage}} \pubyear{2010}
\maketitle

\label{firstpage}

\begin{abstract}
We address the issue of the cosmological bias between matter and galaxy distributions, 
looking at dark-matter haloes as a first step to characterize galaxy clustering.
Starting from the linear density field at high redshift,
we follow the centre of mass trajectory of the material that will form each halo at late times (proto-halo). 
We adopt a fluid-like description for the evolution of perturbations in the proto-halo distribution, 
which is coupled to the matter density field via gravity.
We present analytical solutions for the density and velocity fields, in the context of renormalized perturbation
theory. We start from the linear solution, then compute one-loop corrections for the propagator and the power spectrum.
Finally we analytically resum the propagator and we use a suitable extension of the time-renormalization-group method
\citep{b1} to resum the power spectrum. 
For halo masses $M<10^{14} h^{-1} M_{\odot}$ our results at $z=0$ are in good agreement with N-body simulations.
Our model is able to predict the halo-matter cross spectrum with an accuracy of $5$ per cent up to
$k\approx 0.1\ h$ Mpc$^{-1}$ approaching the requirements of future galaxy redshift surveys.
\end{abstract}


\section{Introduction}
Redshift surveys have shown that the clustering properties of
galaxies strongly depend on their luminosity, color and
morphological (or spectral) type
(e.g. \citealt{b12}; \citealt{b13}).
This indicates that galaxies do not perfectly trace the distribution of
the underlying dark matter, a phenomenon commonly referred to as `galaxy
biasing'.
Its origin lies in the details of the galaxy formation process
which is shaped by the interplay between complex hydrodynamical and
radiative processes together with the dark-matter driven formation of the
large-scale structure.

Attempts to infer cosmological parameters from galaxy clustering studies
are severely hampered by galaxy biasing.
A number of theoretical arguments and the outcome of numerical simulations
both suggest that, on sufficiently large scales,
the power spectra of galaxies and matter should be proportional to each
other:
$P_{\rm g}=b_1^2\, P_{\rm m}$ where the linear bias factor
$b_1$ depends on galaxy type but is generally scale independent
(e.g. \citealt{b14}; \citealt{b15}).
Similarly, to model higher-order statistics, such as the galaxy bispectrum,
it is generally assumed that galaxy biasing is a local process such that
$\delta_g=b_1\delta_m+b_2 \delta_m^2/2+\dots$ where $\delta_g$ and
$\delta_m$ are the (smoothed) galaxy and dark-matter density
contrast, respectively \citep{b16}.
However, the reliance of these phenomenological approximations
limits cosmological studies to very large scales whereas
data with better signal-to-noise ratio are already available on much smaller
scales. Moreover, future studies of
baryonic acoustic oscillations (e.g. \citealt{b17}; \citealt{b18})
will require measurements of the matter power-spectrum with percent or even
sub-percent accuracy in order to shed new light on the source of cosmic
acceleration. Understanding and controlling the effects of galaxy biasing
with this precision will be challenging.
All this provides a very strong motivation for developing more accurate
(and physically driven) models of galaxy biasing.

A number of authors have used the power spectrum statistics to explore
the scale dependence of galaxy biasing based on numerical simulations
(\citealt{b19}; \citealt{b20}; \citealt{b21}; \citealt{b22}; \citealt{b23}; \citealt{b24})
or analytical calculations (\citealt{b25}; \citealt{b26}; \citealt{b27}; \citealt{b8})
stemming from either perturbation theory or
the halo model for the large-scale structure
(see \citealt{b28} for a review).
The general picture is that galaxy biasing is expected to be scale dependent
(i.e. $P_{\rm g}(k)=b(k)^2\, P_{\rm m}(k)$) and the functional form of $b(k)$
can sensibly depend on the selected tracer of the large-scale structure.

Since galaxies are expected to form within dark-matter haloes,
understanding the clustering properties of the haloes is
a key step to accurately model galaxy biasing.
This is a much simpler problem,
considering that dark-matter haloes form under the sole action of gravity.
It is in fact expected that long-wavelength density fluctuations
modulate halo formation by modifying the collapse time of localized
short-wavelength density peaks
(\citealt{b29}; \citealt{b30}).
This argument (known as the peak-background split) predicts that, on
large scales, the halo overdensity $\delta_{\rm h}=b\, \delta_{\rm m}$
where the bias coefficient $b$ varies with the halo mass \citep{b31}.
The numerical value of the bias coefficient is determined by two different
occurrences: first, haloes form out of highly biased
regions in the linear density field (\citealt{b32}; \citealt{b33})
and, second, they move over time as they are accelerated towards the densest
regions of the large-scale structure by gravity \citep{b31}.
These two phenomena generally go under the name of ``Lagrangian biasing''
and ``Lagrangian to Eulerian passage'', respectively.
\citet{b31} dealt with the second problem by assuming that
long-wavelength density perturbations evolve according to the spherical
top-hat model.
A more sophisticated generalization of the peak-background split
has been presented by \citet{b34} who assumed that also
the large-scale motion of the density ``peaks'' is fully determined by
the long-wavelength component of the density field.
Since the halo population and the matter feel the same large-scale
gravitational potential, their density fluctuations are strongly coupled
and their time evolution must be solved simultaneously.
This makes the process of halo biasing non-linear and non-local
even if one starts from a linear and local Lagrangian biasing scheme
(\citealt{b34}; \citealt{b35}).
The bispectrum can be used to test this model against the standard
Eulerian local biasing scheme \citep{b36}.

In this paper, we present a novel and very promising approach to model the
clustering of dark matter haloes. Adopting the formalism by \citet{b34}
combined with a non-local Lagrangian biasing scheme for the haloes
\citep{b4},
we simultaneously follow the growth of perturbations in the matter and in the
halo distribution over cosmic time.
We present perturbative solutions for the corresponding overdensity and
velocity fields and we are able to resum the perturbative series
in the limit of large wavenumbers.
Moreover, we write down a system of equations for the power
spectra $P_{\rm m}$ and $P_{\rm h}$
using the time-renormalization-group (TRG) approach by \citet{b1}
and numerically integrate them. Our results are in excellent agreement
with the output of a high-resolution N-body simulation, showing an improvement over linear theory, and
we are able to predict the matter-halo cross spectrum with a precision within $5$ per cent for
$k< 0.15\ h$ Mpc$^{-1}$.

Related work has been very recently presented by \citet{b42} who 
computed the two-point correlation function of linear density peaks
and followed its time evolution assuming that peaks move according to the
Zel'dovich approximation. For massive haloes this
results in a scale-dependent bias (with variations of $\sim 5$ per cent) 
on the scales relevant for baryonic-oscillation studies.
Contrary to their approach, we do not deal with a point process but
describe large-scale fluctuations in the distribution of dark-matter haloes 
as perturbations in a continuous fluid. On the other hand, we account for
the full gravitational motion of the haloes and do not rely on simplified dynamical models
as like as the Zel'dovich approximation.

The structure of our paper is as follows. In Section \ref{mod} we present
our model for the joint evolution of the matter and halo power spectra. 
The initial conditions for our evolutionary equations are discussed in Section \ref{init}.
The solution of the linearized equations is presented in Section \ref{lin} where we also 
quantify the importance of the halo velocity bias.
Using a perturbative technique,
in Section \ref{analit} we compute
analytic solutions for the propagator of perturbations (the two-time
correlator). We derive 1-loop corrections
and, in the limit of large wavenumbers, the fully resummed propagator.
The discussion in Sections \ref{comp1} and \ref{comp2} is very technical and the less experienced
readers can safely skip it without compromising understanding of the
remainder.
In Section \ref{num}, we numerically integrate the full equations for the
evolution of halo and matter power spectra in the TRG formalism.
We then compare the results
against the outcome of a high-resolution N-body simulation.
Finally, in Section \ref{con} we conclude.

\section{The model}
\label{mod}
\subsection{Dynamics of gravitational instability}
The large-scale structure observed today in the universe is believed to be the
result of gravitational amplification of primordial fluctuations, caused by
the interaction among cold dark matter (CDM) particles. If we denote
$\delta_m$ as the matter density contrast and ${\bf v}$ as the velocity, the Eulerian dynamics of a
system of such particles, which interact only via gravity, is ruled by a set
of three equations (continuity, Euler and Poisson) that in a $\Lambda$CDM model reads:
\begin{eqnarray}
&&\frac{\partial\,\delta_m}{\partial\,\tau}+
{\bf \nabla}\cdot\left[(1+\delta_m) {\bf v} \right]=0\,,\nonumber\\
&& \frac{\partial\,{\bf v}}{\partial\,\tau}+{\cal H}\,{\bf v}\, + ( {\bf v} 
\cdot {\bf \nabla})  {\bf v}= - {\bf \nabla} \phi\,,\nonumber\\
&&\nabla^2 \phi = \frac{3}{2}\,\,{\cal H}^2  \,\Omega_m \, \delta_m \,,
\label{a}
\end{eqnarray}
where $\tau$ is the conformal time.
If we define the velocity divergence $\theta(\bx,\,\tau) = 
\nabla \cdot {\bf v} (\bx, \,\tau)$ and switch to Fourier space, the equations
in (\ref{a}) become:
\begin{eqnarray}
&&\frac{\partial\,\delta_m({\bf k}, \tau)}{\partial\,\tau}+\theta({\bf k}, 
\tau) \nonumber\\
&& + \int d^3\bq\, d^3\bp \,\delta_D({\bf k}-\bq-\bp)
 \alpha(\bq,\bp)\theta(\bq, \tau)\delta_m(\bp, \tau)=0\,,\nonumber\\
&&\frac{\partial\,\theta({\bf k}, \tau)}{\partial\,\tau}+
{\cal H}\,\theta({\bf k}, \tau)  +\frac{3}{2} {\cal H}^2 \,\Omega_m(\tau)
\delta_m({\bf k}, \tau)\nonumber\\
&& +\int d^3\bq \,d^3\bp \,\delta_D({\bf k}-\bq-\bp) 
\beta(\bq,\bp)\theta(\bq, \tau)\theta(\bp, \tau) = 0 \,,\nonumber\\
&& \nonumber\\
\label{b}
\end{eqnarray}
where
\begin{equation}
\alpha(\bq,\bp )= \frac{(\bp + \bq) \cdot \bq}{q^2}\,,\quad \quad
\beta(\bq,\bp ) = \frac{(\bp + \bq)^2 \,
\bp \cdot \bq}{2 \,p ^2 q^2}\,,
\label{c}
\end{equation}
Equations (\ref{b}) can be written in a compact form if we define a new time 
variable $\eta\equiv\ln(D_+/D_{+in})$, being
$D_{+in}$ the growth factor at an early epoch, and a doublet $\varphi_a$ ($a=1,2$)
\begin{equation}
\left(\begin{array}{c}
\varphi_1 ( {\bf k}, \eta)\\
\varphi_2 ( {\bf k}, \eta)  
\end{array}\right)
\equiv 
e^{-\eta} \left( \begin{array}{c}
\delta_m  ( {\bf k}, \eta) \\
-\theta  ( {\bf k}, \eta)/({\cal H}f_+)
\end{array}
\right)\,,
\label{d}
\end{equation}
with $f_+\equiv d\ln D_+/d \ln a$.
The velocity divergence is scaled such that it has the same dimension of the density contrast
and in the linear regime $-\theta  ( {\bf k}, \eta)/({\cal H}f_+)\approx\delta_m  ( {\bf k}, \eta)$, i.e. $\varphi_1 ( {\bf k}, 0)=\varphi_2 ( {\bf k}, 0)$.
The system is therefore 
\begin{eqnarray}
&&\partial_\eta\,\varphi_a({\bf k}, \eta)= -\Omega_{ab}(\eta )
\varphi_b({\bf k}, \eta) \nonumber\\
&&\qquad \qquad\quad+ e^\eta 
\gamma_{abc}({\bf k},\,-{\bf p},\,-{\bf q})  
\varphi_b({\bf p}, \eta )\,\varphi_c({\bf q}, \eta ),
\label{e}
\end{eqnarray}
where sum over repeated indices and integration over repeated momenta are
understood.
The vertex function $\gamma_{abc}({\bf k},{\bf p},{\bf q}) $ ($a,b,c,=1,2$)
has only three non-vanishing elements
\begin{eqnarray}
&&\gamma_{121}({\bf k},\,{\bf p},\,{\bf q}) = 
\frac{1}{2} \,\delta_D ({\bf k}+{\bf p}+{\bf q})\, 
\alpha(\bp,\bq)\,,\nonumber\\
&&\gamma_{222}({\bf k},\,{\bf p},\,{\bf q}) = 
\delta_D ({\bf k}+{\bf p}+{\bf q})\, \beta(\bp,\bq)\,,
\label{f}
\end{eqnarray}
and $\gamma_{112}({\bf k},\,{\bf q},\,{\bf p})=\gamma_{121}({\bf k},\,{\bf
  p},\,{\bf q})$, with $\delta_D$ the Dirac-delta distribution.
All the information about the cosmological model is contained in the matrix 
\begin{equation}
{\bf \Omega} (\eta ) = \left(\begin{array}{cc}
\displaystyle 1 & \displaystyle -1\\
\displaystyle -\frac{3 \Omega_m}{2f_+^2}  &\displaystyle \frac{3 \Omega_m}{2f_+^2} \end{array}
\right)\,,
\label{g}
\end{equation}
that, in the following, will be considered as a constant matrix, approximating
$\frac{\Omega_m}{f_+^2}\approx 1$. This ratio is indeed very close to unity for most of the history
of the Universe. Making this approximation,
one is modifying the behavior of the decaying mode, while the growing one is left
unaltered. It has been shown in \cite{b1} that it affects the matter power spectrum at $z=0$ at a
less than percent level up to $k\approx 0.3 \,h$ Mpc$^{-1}$.\\

The power spectrum, defined by an ensemble average, in this notation is a $2\times2$ matrix
\begin{equation}
\langle \varphi_a({\bf k};\,\eta) \varphi_b({\bf q};\,\eta)\rangle \equiv
\delta_D({\bf k + q}) P_{ab}(k;\,\eta)\,,
\label{h}
\end{equation}
and the bispectrum, defined by
\begin{eqnarray}
&&\langle \varphi_a({\bf k};\,\eta) \varphi_b({\bf q};\,\eta)\varphi_c({\bf
  p};\,\eta)\rangle\nonumber\\
&& \equiv \delta_D({\bf k + q+p})
 B_{abc}({\bf k},\,{\bf q},\,{\bf p};\,\eta)\,,
\label{i}
\end{eqnarray}
has 8 components.
In the following, we will also consider a different-time two-point correlator, defined as
\begin{equation}
\langle \vp_a({\bf k},\eta_a) \vp_b({\bf q},\eta_b)\rangle \equiv
\delta_D({\bf k + q}) P_{ab}(k;\,\eta_a,\eta_b)\,,
\label{2pc}
\end{equation}
which obviously coincides with (\ref{h}) for $\eta_a=\eta_b$.

\subsection{The distribution of dark-matter haloes}
Let us consider a set of dark-matter haloes identified at a given redshift
$z_{\rm id}$ according to some predefined criterion.
The material that forms the haloes can be traced back to its initial location
in the linear overdensity field at $z\to \infty$.
We dub each of these regions as a proto-halo.
In other words, a proto-halo is the Lagrangian region of space that will
collapse to form a halo at redshift $z_{\rm id}$.
N-body simulations show
that nearly all proto-haloes are simply connected
\citep{b10} even though this property is not
key to our analysis.

Let us now follow the evolution of a proto-halo over cosmic time in Eulerian
space. Basically its
shape and topology will be distorted (proto-haloes will first
fragment into smaller substructures that will later merge to form the final
halo) and its overall volume will be compressed
while its centre of mass will move along a given trajectory determined
by the mass density field via gravity.
We focus our analysis onto this motion that connects the Lagrangian
position of the proto-halo with the Eulerian location of the final halo.

On scales much larger than the characteristic size of (and separation between) 
the proto-haloes,
the density fluctuations traced by the centre-of-mass trajectories
can be described with a continuous overdensity field
$\delta_{\rm h}({\bf x},\tau | z_{\rm id})$. Note that while $\tau$ labels
conformal time along the trajectories,
$z_{\rm id}$ is just a tag that identifies the halo population.
Unlike real haloes that undergo merging, 
by construction proto-haloes always preserve their
identity. Their total number is therefore
conserved over time and we can write a continuity equation for their
number density:
\begin{equation}
\frac{\partial \delta_{\rm h}}{\partial \tau}+ \nabla \cdot [(1+\delta_{\rm h})
{\bf v}_{\rm h}]=0\;.
\label{j}
\end{equation}
Here the proto-halo density and velocity fields should be intended as  
coarse grained on a scale of a few times the mean inter-halo  
separation (so as to suppress discreteness effects as proto-haloes are  
individually separate units).
Strictly speaking, the smoothed velocity field does not obey
the Euler-Poisson system in equation (\ref{a}) due to the presence
of the non-linear term $({\bf v}\cdot \nabla) {\bf v}$.
In fact, the coarse graining procedure introduces new terms in the  
fluid equations generated by the degrees of freedom one has integrated  
out, namely: a velocity dispersion term and a correction to the mean- 
field gravitational acceleration due to density fluctuations
on scales smaller than the smoothing radius (\citealt{b46}).
On the other hand, it is reasonable to assume that 
the large-scale motion of the proto-haloes is generated by density fluctuations
with wavelength larger than the characteristic halo size and is not  
influenced by perturbations with shorter wavelength.
The very same assumption of neglecting
the coupling to the small scales is routinely done when one writes equation  
(\ref{a}) for the mass density field (see Section 3 in \citealt{b46}) albeit adopting much narrower smoothing kernels than for the  
haloes.

With the same spirit, in what follows, we will ignore the extra  
terms in the fluid equations generated by the coarse graining procedure. This is a working hypothesis which makes
the problem mathematically treatable and whose accuracy
will be tested
by comparing our final results against high-resolution numerical simulations.
We therefore write an Euler equation for the proto-halo fluid velocities
\begin{equation}
\frac{\partial\,{\bf v}_{\rm h}}{\partial\,\tau}+{\cal H}\,{\bf v}_{\rm h}\,
+ ( {\bf v}_{\rm h}
\cdot {\bf \nabla})  {\bf v}_{\rm h}= - {\bf \nabla} \phi\;,
\label{jj}
\end{equation}
where the gravitational potential is the same as in equation (1).
Note that if ${\bf v}_{\rm h}$ matches ${\bf v}$ in the initial conditions 
then it will always do. On the contrary,
any initial velocity bias between proto-halos
and matter will be progressively erased by the gravitational acceleration. 

Thus, given suitable initial conditions for $\delta_{\rm h}$ and
${\bf v}_{\rm h}$ at $\tau \to 0$ (i.e. a prescription for the 
Lagrangian biasing of proto-haloes), we can in principle use
equations (\ref{j}) and (\ref{jj}) to follow the clustering
evolution of the proto-halo population at all times.
We are particularly interested in the solution of the fluid equations 
at the special time $\tau$ that corresponds to $z_{\rm id}$.
In fact this solution has a particular physical meaning as it
gives the density and velocity fields of the actual dark-matter haloes.

\subsection{Growth of matter and halo perturbations}
\label{nle}
The system (\ref{a}) is now extended by the inclusion of eq. (\ref{j}) and eq. (\ref{jj}).
We define a quadruplet $\varphi_a$ ($a=1,2,3,4$) 
\begin{equation}
\left(\begin{array}{c}
\varphi_1 ( {\bf k}, \eta)\\
\varphi_2 ( {\bf k}, \eta)\\
\varphi_3 ( {\bf k}, \eta)\\ 
\varphi_4 ( {\bf k}, \eta)\\ 
\end{array}\right)
\equiv 
e^{-\eta} \left( \begin{array}{c}
\delta_m  ( {\bf k}, \eta) \\
-\theta  ( {\bf k}, \eta)/({\cal H}f_+)\\
\delta_h  ( {\bf k}, \eta)\\
-\theta_h  ( {\bf k}, \eta)/({\cal H}f_+)\\
\end{array}
\right)\,,
\label{new1}
\end{equation}
in such a way that eq. (\ref{e}) still holds, but with indices running from 1 to 4.
There are three more non-vanishing elements of the vertex $\gamma_{343}({\bf k},\,{\bf p},\,{\bf q})=\gamma_{334}({\bf k},\,{\bf q},\,{\bf p})=\gamma_{121}({\bf k},\,{\bf
  p},\,{\bf q})$ and $\gamma_{444}({\bf k},\,{\bf q},\,{\bf p})=\gamma_{222}({\bf k},\,{\bf
  p},\,{\bf q})$, and the $4\times4$ ${\bf \Omega}$ matrix is
\begin{equation}
{\bf \Omega} = \left(\begin{array}{cccc}
\displaystyle 1 & \displaystyle -1 & 0 & 0\\
\displaystyle -\frac{3}{2}  & \displaystyle \frac{3}{2} & 0 & 0 \\
0 &0 &\displaystyle 1 & \displaystyle -1\\
\displaystyle -\frac{3}{2}  & 0 & 0 & \displaystyle \frac{3}{2}\end{array}
\right)\,.
\label{new2}
\end{equation}

From the definitions (\ref{h}) and (\ref{i}), with $a=1,2,3,4$, we get a $4~\times~4$
matrix for power spectrum; in the
following, we will focus on the matter power spectrum
$P_{11}$ and the matter-halo cross spectrum $P_{13}$.

\section{Initial conditions}
\label{init}
In the previous section, we have presented a model that  
describes the non-linear evolution of the matter and halo density fields.
Given suitable initial conditions, the formal equations we have introduced 
can be integrated numerically so that to compute
the perturbative propagators and the TRG-evolved power spectra.
The choice of the initial conditions therefore plays a very important role
in our theory and will be the subject of this section.

\subsection{N-body simulation}
To gain insight into the properties of proto-haloes 
(and, later, to test our results at $z=0$), 
we use one high-resolution N-body simulation extracted from the suite presented
by Pillepich et al. (2010).
This contains $1024^3$ dark-matter particles within a periodic cubic box 
with a side of $L_{box}=1200 h^{-1}$ Mpc and follows the formation of structure
in a $\Lambda$CDM model with Gaussian initial conditions and
cosmological parameters: 
$h=0.701$, $\sigma_8=0.817$,
$n_s=0.96$, $\Omega_m=0.279$, $\Omega_b=0.0462$ and
$\Omega_{\Lambda}=0.721$.

We identify dark-matter haloes at $z=0$ using 
the friends-of-friends algorithm with
a linking length equal to 0.2 times the mean interparticle distance.
We only consider haloes containing more than 100 particles  
(i.e. with mass $M>1.24\cdot10^{13} h^{-1} M_{\odot}$) and we
split them into four mass bins to keep track of their different clustering
properties.
The corresponding mass ranges and the total number of haloes in each bin
are given in Table \ref{tab:1}, along with an estimate of the highest wavevector up to which the fluid approximation for haloes holds. This value is determined by the number of haloes we require to be in a volume element to consider them as a fluid, and we set this number to 30. On smaller scales, our assumption breaks down, therefore we will look at results in the specified range, that, of course, decreases as the mass of the haloes increases. In the plots that will be shown in Section \ref{num} the limit to which we can trust our model will be represented by vertical black dotted lines.

Proto-haloes are identified by tracing the positions of the particles forming 
a halo at $z=0$ back to the linear density field. The centre of mass  
of each proto-halo is used as a proxy for its spatial location. Similarly,
the mass weighted linear velocity gives the proto-halo velocity.

Halo and proto-halo density and momentum fields are computed  
with the cloud-in-cell grid assignement using a $512^3$ mesh.
Velocity fields are obtained by taking the ratio of the momentum and density
distributions (preventively smoothed to preclude the existence of empty cells)
as shown in \citet{b37}. 

Power spectra have been computed using FFT. In order to avoid uncertain
shot-noise corrections for the haloes, we only consider their cross spectra
with the matter density field.

\subsection{Lagrangian halo bias}
Concerning the matter density, the initial conditions are given by linear
theory which directly 
follows from the adopted cosmological model (transfer function)
and the statistics of primordial perturbations (spectral index, Gaussianity).
On the other hand, for the dark-matter halos, 
we can follow two different approaches: 
(i) extract the relevant information directly from the simulation
or (ii) use a model for the Lagrangian bias of the halos.
The latter option offers a number of advantages. 
First, it allows us to make general predictions independently of 
the simulation specifics.
Second, it allows us to include halo bispectra in our formalism 
(while it would be extremely demanding and time consuming to compute all 
possible triangular configurations from the simulation).
For these reasons we will present below a model for the bias of the 
proto-haloes.
Note, however, that any lack of accuracy of the adopted Lagrangian biasing
scheme will propagate through the time evolution of our model
and contribute to the imprecision of its final results.
Therefore, in order to test the accuracy of our evolutionary equations alone,
we will also extract initial conditions directly from the simulation and
compare the corresponding outcome of the evolution model with the statistics
of the simulated haloes at $z=0$.

Let us consider the overdensity of proto-haloes in Lagrangian space 
$\delta_h({\bf q})$ and the corresponding mass-density fluctuation
$\delta({\bf q})$. We assume that their Fourier transforms are linked
by the expression:
\begin{equation}
\delta_h({\bf k})=(b_1+b_2\cdot k^2)\, \delta_m({\bf k}),
\label{w}
\end{equation}
which corresponds to a non-local relation in real space.
This form was first proposed by \citet{b4} and describes the clustering of
linear density peaks \citep{b9}. 
In this case, the initial conditions for
$P_{33}$ and $P_{13}$ are:
\begin{eqnarray}
P_{33}(k)&=&(b_1+b_2 k^2)^2 P_{11}(k) \mathrm{e}^{-k^2R^2}\,,\nonumber\\
P_{13}(k)&=&(b_1+b_2 k^2) P_{11}(k)\mathrm{e}^{-k^2R^2/2}\,,
\label{x}
\end{eqnarray}
where the exponential functions accounts for the finite size of the density
peaks corresponding to a given halo mass.
%
We find that the expression above accurately describes the cross-spectrum
of proto-haloes and matter in our N-body simulation when $b_1$, $b_2$ and $R$
are treated as fitting parameters (see \citealt{b41} for further details).
\footnote{Note that, adopting a local-bias model, $\delta_h({\bf q})=b_1\cdot
\delta_m({\bf q})+b_2\cdot \delta_m^2({\bf q})$, provides a worse fit to
simulation results. 
In this case, the model $P_{13}$ lacks power for all but the smallest 
wavenumbers.}
This is not surprising as \citet{b38} have shown that most
proto-haloes include a density peak of the corresponding mass scale
within their Lagrangian volume. 
In Table \ref{tab:1} we quote, for each halo-mass bin, 
the parameters $b_1$, $b_2$ and $R$ that best fit the simulation data
using eq. (\ref{x}) where
the linear matter power spectrum $P_{11}$ is computed using the CAMB 
online tool \citet{b7}. 

Whereas Gaussianity is a good approximation for the linear matter distribution,
fluctuations in the halo counts are non-Gaussian even in the initial conditions.
We can quantify their level of non-Gaussianity in terms of their auto and
cross bispectra that,
using equation (\ref{w}), can all be reduced to one of the following
forms:
\begin{eqnarray}
B_{333}({\bf k_1},{\bf k_2},{\bf k_3})\phantom{i}&=&(b_1+b_2 k_1^2)(b_1+b_2 k_2^2)(b_1+b_2
k_3^2)\nonumber\\
&&B_m({\bf k_1},{\bf k_2},{\bf k_3})\,,\nonumber\\
B_{133\phantom{i}}({\bf k_1},{\bf k_2},{\bf k_3})&=&(b_1+b_2 k_2^2)(b_1+b_2 k_3^2)B_m({\bf k_1},{\bf k_2},{\bf k_3})\,,\nonumber\\
B_{113}({\bf k_1},{\bf k_2},{\bf k_3})&=&(b_1+b_2 k_3^2)B_m({\bf k_1},{\bf k_2},{\bf k_3})\,,
\label{y}
\end{eqnarray}
where
the matter bispectrum $B_m({\bf k_1},{\bf k_2},{\bf k_3})$ is computed 
using the tree-level expression of the standard perturbation
theory,
\begin{eqnarray}
B_m({\bf k_1},{\bf k_2},{\bf k_3})&=&2\left[\frac{1}{2} +
  \frac{1}{2}\left(\frac{k_1}{k_2}+\frac{k_2}{k_1}\right)\mu_{12}+\frac{1}{2}\mu_{12}^2\right]\nonumber\\
&& P_{11}(k_1,z_{in})P_{11}(k_2,z_{in})+{\rm cycl.}\,,
\end{eqnarray}
with $\mu_{12}\equiv\frac{{\bf k}_1\cdot{\bf k}_2}{k_1k_2}$.\\

In order to solve our evolutionary equations, we need to know also the linear velocity field of proto-haloes.
In principle proto-haloes might not move with the same velocity as matter at the same location
(at the very least they should match the mass velocity smoothed on the Lagrangian halo size). 
We model this effect assuming that proto-haloes are indeed related to linear density peaks which,
as discussed above, gives a good description of their clustering properties.
In particular we follow \citet{b40} who proposed a model for the peak velocities,
which in Fourier space assumes the form
\beq
\theta_h(\bk)=\left(1-\frac{\sigma_0^2}{\sigma_1^2}k^2\right) e^{-k^2R^2/2} \theta_m(\bk)\equiv b_v(\bk)\theta_m(\bk)\,,
\label{vbias}
\eeq
with $b_V$ the scale-dependent ``velocity bias'' and 
$\sigma_n$ ($n=0,1$) being the spectral moments of the matter power spectrum
defined as
\beq
\sigma_n^2 = \frac{1}{2 \pi^2} \int_0^{\infty} \mathrm{d}k\,k^{2(n+1)}\,P_{11}(k)\,e^{-k^2R^2}\,.
\eeq
In the equations above
a Gaussian smoothing window of size $R$ has been adopted to coarse grain the matter density
and identify the peaks.
Once the proto-halo mass is linked to $R$ through $M=(2\pi)^{3/2}\,\bar{\rho}\,R^3$ (with
$\bar{\rho}$ the mean comoving density of matter),
this model has no free parameters, since the spectral moments are completely
defined by $P_{11}(k)$. 
It follows that the initial conditions for $P_{24}$ and $P_{44}$ are
\begin{eqnarray}
P_{44}(k)&=&b_v(k)^2\, P_{22}(k)\,,\nonumber\\
P_{24}(k)&=&b_v(k)\, P_{22}(k)\,,
\label{xx}
\end{eqnarray}
and these expressions are in very good agreement with the spectra computed from the N-body simulation 
(\citealt{b41}). The corresponding values of $\sigma_0^2/\sigma_1^2$ are listed in 
Table \ref{tab:1} as a function of halo mass. Note that the velocity bias becomes more and more important
on large scales with increasing the halo mass.

\begin{table*}
\begin{tabular}{|c|c|c|c|c|c|c|c|c|}
\hline
Bin & Mass range & $\#$ haloes & $k_{max}$ & $b_1$ & $b_2$ & $R$ & $\sigma_0^2/\sigma_1^2$ \\
 & ($10^{13} M_{\odot}/h$) &  & ($\mathrm{Mpc}^{-1} \,h$)&  & ($\mathrm{Mpc}^2 \,h^{-2}$) & ($\mathrm{Mpc} \,h^{-1}$) & ($\mathrm{Mpc}^2 \,h^{-2}$)\\ \hline
Bin 1 &  $1.24-1.8$ & $202948$ & $0.24$ &  $7.28 \pm 0.38$ & $422 \pm 102$ & $2.7 \pm 0.8 $ & $9.1$\\ 
Bin 2 &  $1.8-3.4$ & $211305$ & $0.24$ & $14.2 \pm 0.4$ & $356 \pm 84$ & $2.1 \pm 0.7 $ & $12.3$\\ 
Bin 3 &  $3.4-10$ & $150105$ & $0.22$ &  $25.9 \pm 0.4$ & $708 \pm 103$ & $2.9 \pm 0.4 $ & $20.5$\\ 
Bin 4 &  $>10$ & $48985$ & $0.15$ &  $66.2 \pm 1.3$ & $1025 \pm 401$ & $3.5 \pm 0.8 $ & $50.9$\\ \hline
\end{tabular}
\caption{Mass range and number of the haloes in the four bins.}
\label{tab:1}
\end{table*}

\section{Linear theory}
\label{lin}
The lowest order approximation to the perturbation equation (\ref{e}) consists in setting $\gamma_{abc}=0$. In this limit, the evolution of the field from the initial time $\eta=0$ to a generic $\eta$ is given by
\beq
\vp_a(\bk;\eta)=g_{ab}(\eta)\vp_b(\bk;0)\,,
\label{new3}
\eeq
where $g_{ab}(\eta)$ is the {\em linear propagator}, defined by the equation
\beq
\left(\delta_{ab}  \partial_\eta + \Omega_{ab}\right) g_{bc}(\eta) = \delta_{ac} \delta_{D}(\eta)\,,
\label{new4}
\eeq
with $\delta_{ab}$ the Kronecker delta. Solving eq.~(\ref{new4}) with causal boundary conditions ($g_{ab}(\eta)=0$ for $\eta<0$, see e.g. \citealt{b5}) one gets
\begin{eqnarray}
g_{ab}(\eta) &=& \left[\left( \begin{array}{cccc} 3/5 & 2/5 & 0 & 0\\ 3/5 & 2/5
    & 0 & 0\\ 3/5 & 2/5 & 0 & 0\\ 3/5 & 2/5 & 0 & 0\end{array} \right) \right.\nonumber\\
&& + e^{-5/2\eta}\left( \begin{array}{cccc} 2/5 & -2/5 & 0 & 0\\ -3/5 &
    3/5 & 0 & 0\\ 2/5 & -2/5 & 0 & 0 \\-3/5 &
    3/5 & 0 & 0\end{array} \right)\nonumber\\
&& + e^{-3/2\eta}\left( \begin{array}{cccc} 0 & 0 & 0 & 0\\ 0 &
    0 & 0 & 0\\ 0 & 2 & 0 & -2 \\0 &
    -1 & 0 & 1\end{array} \right)\nonumber\\    
&& \left.+ e^{-\eta}\left( \begin{array}{cccc} 0 & 0 & 0 & 0
    \\ 0 & 0 & 0 & 0\\ -1 & -2 & 1 & 2 \\0 & 0 & 0 & 0\end{array} \right)\right]\,\theta(\eta)\,,
\label{new5}
\end{eqnarray}
with $\theta(\eta)$ Heavyside's step function.
Notice that $g_{ab}(\eta)\to\delta_{ab}$ as $\eta\to0^+$.
The first and second contributions represent the standard growing and decaying modes, respectively ~\citep{b5}. The third and fourth contributions represent two new modes, decaying respectively as $e^{-3\eta/2}$ and $e^{-\eta}$ compared with the growing one. To understand their physical effect we notice that an initial condition of the form
\beq
\vp_a(\bk;0)=\left(\begin{array}{c}
\vp(\bk)\\
\vp(\bk)\\
\vp_h(\bk)\\
\vp_v(\bk)\\
\end{array}
\right)\,,
\label{ic}
\eeq
evolves (using eq.~(\ref{new3}))  into $\vp_a(\bk;\eta)$ given by
\beq
\left(\begin{array}{c}
\vp\\
\vp\\
\vp+2 e^{-3\eta/2}(\vp-\vp_v)+ e^{-\eta}(-3\vp+\vp_h+2\vp_v)\\
\vp-e^{-3\eta/2}(\vp-\vp_v)\\
\end{array}
\right)\,,
\label{uff}
\eeq
{\it i.e.} both the halo density and velocity fields relax to the matter ones
as $\eta\to\infty$ (but at a different pace).
Also note that in the absence of an initial density bias (i.e. $\phi_3=\phi$) but in the presence
of an initial velocity bias (i.e. $\phi_4\neq\phi$), the linear dynamics quickly generates a transient
density bias that vanishes at late times as $e^{-\eta}-e^{-3\eta/2}$.
The initial power spectrum at $\eta=0$, corresponding to the field configuration in eq.~(\ref{ic}), is $P_{ab}^0(k)$, 
and it evolves forward in time as
\begin{equation}
P_{L,ab}(k;\eta_a,\eta_b) = g_{ac}(\eta_a) g_{bd}(\eta_b) P^0_{cd}(k)\,.
\label{new7}
\end{equation}

\subsection{The importance of velocity bias}
It is interesting to assess the role of the velocity bias in the linear solution previously discussed. 
Assuming $\vp_4=\vp_2$ at all times, the linear propagator for the first three components $\vp_i$ with $i=1,2,3$ becomes
\begin{eqnarray}
g_{ab}(\eta) &=& \left[\left( \begin{array}{ccc} 3/5 & 2/5 & 0 \\ 3/5 & 2/5
    & 0 \\ 3/5 & 2/5 & 0 \end{array} \right) \right.\nonumber\\
&& + e^{-5/2\eta}\left( \begin{array}{ccc} 2/5 & -2/5 & 0 \\ -3/5 &
    3/5 & 0 \\ 2/5 & -2/5 & 0 \end{array} \right)\nonumber\\
&& \left.+ e^{-\eta}\left( \begin{array}{ccc} 0 & 0 & 0
    \\ 0 & 0 & 0 \\ -1 & 0 & 1 \end{array} \right)\right]\,\theta(\eta)\,,
\label{n}
\end{eqnarray}
and the third component of (\ref{uff}) reduces to
\beq
\vp_3(\bk;\eta)=\vp(\bk)+e^{-\eta}(\vp_h(\bk)-\vp(\bk))\,.
\eeq
This expresses the well known {\em linear debiasing} between the halo and matter fields at late times, derived by \citet{b11} 
for tracers that do not undergo merging and move solely under the influence of gravity.

\begin{figure}
\includegraphics[width = 3.5in,keepaspectratio=true]{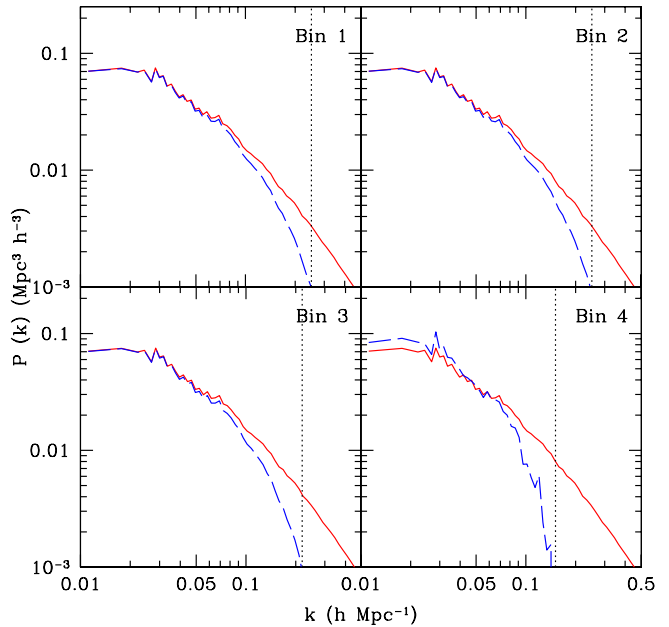}
\caption{A comparison between $P_{14}$ (blue dashed line) and $P_{11}$ (red solid line) in the four bins. 
The vertical black dotted lines represent the limit up to which we expect our fluid approximation for haloes to work.
A smoothing scale of $R=7\,\mathrm{Mpc} /h$ has been used for $P_{14}$. For a fair comparison with $P_{11}$, which is not smoothed, 
$P_{14}$ has been redivided by the smoothing function.}
\label{fig0}
\end{figure}

\begin{figure}
\includegraphics[width = 3.5in,keepaspectratio=true]{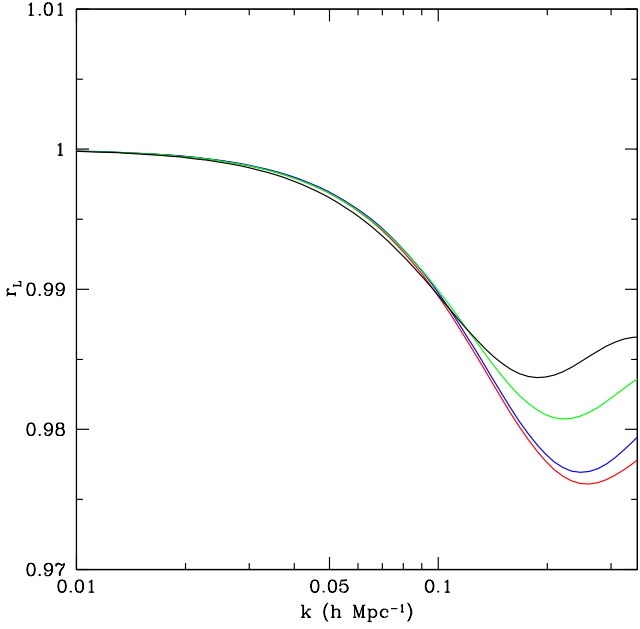}
\caption{The ratio between $P_{L,13}$ and $P_{L,13}^{(3)}$: going from bottom to top, Bin 1 (red), Bin 2 (blue), Bin 3 (green) and Bin 4 (black).}
\label{fig2}
\end{figure}

The corresponding halo-matter cross spectrum is
\beq
P_{L,13}=P^0_{11} + e^{-\eta}(P^0_{13} - P^0_{11})\equiv P_{L,13}^{(3)}\,,
\eeq
while keeping $\vp_4\neq\vp_2$ one gets
\beq
P_{L,13}=P_{L,13}^{(3)} + 2 (P^0_{11}-P^0_{14}) (e^{-3 \eta/2} - e^{-\eta})\,.
\label{vb}
\eeq
In Figure \ref{fig0} we compare $P^0_{11}$ against $P^0_{14}$ extracted from the N-body simulation, for the different mass bins.
While the spectra agree well on very large scales ($k\lesssim 0.05\, h\, \mathrm{Mpc}^{-1}$), they progressively depart for smaller scales.
This is in line with the model introduced in Section 3.
Note that the last term in eq. (\ref{vb}) vanishes in the initial conditions, reaches a minimum for $\eta\approx0.8$ and it is suppressed at late times.
We quantify its amplitude at $z=0$ in Figure \ref{fig2}, where we plot the ratio $r_L=P_{L,13}/P_{L,13}^{(3)}$
which ranges between $0-3$ per cent, depending on halo mass and scale.
This suggests that the effect of the velocity bias on the halo-matter cross spectrum is small for low redshifts.
\section{Analytical treatment of non-linearities}
\label{analit}
In this section we deal with the non-linear evolution of the matter-halo system.
We first compute 1-loop corrections for the propagator 
and then perform the corresponding large-k resummation to all perturbative orders 
thus extending the results presented by \citet{b5} for the matter density field.
Finally, we compute non-linear power spectra using the TRG approach.
For simplicity we only consider the case with no velocity bias, which we have demonstrated to be accurate 
(at least in the linear regime) at low redshifts, where current observations are available.
The $3\times3$ ${\bf \Omega}$ matrix for this case is
\begin{equation}
{\bf \Omega} = \left(\begin{array}{ccc}
\displaystyle 1 & \displaystyle -1 & 0 \\
\displaystyle -\frac{3}{2}  & \displaystyle \frac{3}{2} & 0 \\
0 &\displaystyle -1 & \displaystyle 1\\\end{array}
\right)\,.
\label{l}
\end{equation}
\subsection{1-loop perturbation theory}
\label{comp1}
The 1-loop correction to the linear propagator \citep{b5} is given by
\begin{eqnarray}
\Delta g_{ab}(k;\eta) &=& 4 \int_{0}^{\eta} ds_1 \int_{0}^{s_1} ds_2 \int d^3q\, e^{s_1+s_2}\nonumber\\ 
&& g_{ac}(\eta-s_1) \gamma_{cie}({\bf k},\,{\bf -q},\,{\bf q-k})g_{ef}(s_1-s_2)\nonumber\\ 
&&\gamma_{fhd}({\bf k-q},\,{\bf q},\,{\bf -k})g_{db}(s_2)P_{L,ih}(q,s_1,s_2)\,,\nonumber\\
&&
\label{r}
\end{eqnarray}
(see appendix A for its explicit expressions).

The 1-loop correction to the two-point correlator (\ref{2pc}) is given by the sum of two contributions
\beq
\Delta P_{ab}^{I\phantom{ab}}(k;\eta_a,\eta_b)+\Delta P_{ab}^{II\phantom{ab}}(k;\eta_a,\eta_b)\,,
\eeq
which are also known in the literature as  ``$P_{13}$" and ``$P_{22}$", respectively \footnote{They should not be confused with the $P_{13}$ and $P_{22}$ of our notation!}. They are given by
\beqra
 \Delta P_{ab}^{I\phantom{ab}}(k;\eta_a,\eta_b) & = &\Delta g_{ac}(k;\eta_a) \,g_{bd}(k;\eta_b) \, P_{cd}^0(k) \nonumber\\
&& + (a\leftrightarrow b)\,,
\eeqra
\beqra
\Delta P_{ab}^{II\phantom{ab}}(k;\eta_a,\eta_b) &=&  \int_0^{\eta_a} ds_a \int_0^{\eta_b} ds_b \, \Phi_{cd}(k;s_a,s_b) \nonumber\\
&& g_{ac}(k;\eta_a-s_a) \,g_{bd}(k;\eta_b-s_b)\,,
\eeqra
with
\beqra
 \Phi_{cd}(k;s_a,s_b) &=& 2 e^{s_a+s_b} \int d^3q  \nonumber\\
 &&\gamma_{cei}({\bf k},\,{\bf -q},\,{\bf q-k})P_{L,ef}(q,s_a,s_b)\nonumber\\ 
&& P_{L,ih}(|{\bf k}-{\bf q}|,s_a,s_b)\gamma_{dfh}({\bf -k},\,{\bf q},\,{\bf k-q})\,. \nonumber\\
\label{phi}
\eeqra
The explicit expressions for $\Delta P_{ab}^{II}$ are given in appendix A.
\subsection{Large-k resummation for the propagator}
\label{comp2}
In the large-$k$ limit, 
the 1-loop correction for the propagator, eq.~(\ref{r}), grows as $k^2$, and eventually dominates over the ($k$-independent) linear propagator. Taking into account higher orders, the situation gets even worse. The  2-loop
correction grows as $k^4$, the 3-loop as $k^6$, and so on. This a
manifestation of the perturbative expansion breakdown in cosmological PT,
which appears not only in the computation of the propagator, but also of the
power spectrum, the bispectrum, and so on. However, for the case of the
propagator, it was shown in \citet{b6} that the leading order corrections in
the large $k$ and large $\eta$ limit can be resummed at all orders in
perturbation theory, giving a well-behaved propagator.\\
The propagator $ {\bf G_{ac}(k;\eta)}$ connects the initial correlators with the cross-correlations between final and initial field configurations, 
\beq
 P_{ab}(k; \eta,0) = G_{ac}(k;\eta) P^0_{cb}(k) +\Delta P_{ab}^{NG}(k;\eta,0)\,,
 \label{ct}
 \eeq
where the last term at the rhs comes from the initial non-Gaussianity of the matter and halo fields. At leading order, it is given by (see appendix \ref{ang})
\begin{eqnarray}
 &&\Delta P_{ab}^{NG}(k;\eta,0) = \int_0^\eta ds \,e^s g_{ac}(\eta-s) g_{df}(s)g_{eg}(s)  \,\nonumber\\
 && \times \int d^3 q\, \gamma_{cde}({\bf k}, {\bf -q}, {\bf q-k}) B_{fgb}({\bf q}, {\bf k-q},{\bf -k})\,,
\label{eng}
\end{eqnarray}
where $B_{abc}$ is the initial bispectrum at $\eta=0$ ($z=z_{in}$) (see also \citealt{b45}).
In Section \ref{propres} we will use eq.~(\ref{ct}) to assess the validity of different approximation schemes for the propagator.
In the large-$k$ limit, $G$ decays as 
\begin{equation}
G_{ab}(k;\eta)=g_{ab}(\eta) \exp\left(-\frac{k^2\sigma^2e^{2\eta}}{2} \right)\,, 
\label{s}
\end{equation}
 with
\begin{equation}
\sigma^2=\frac{1}{3}\int d^3q\, \frac{P^0(q)}{q^2}\,.
\label{t}
\end{equation}
Therefore, at least in the case of the propagator, the bad ultraviolet behavior is just an artifact of the perturbative expansion, which, at any finite order, completely misses the nice --and physically motivated-- Gaussian decay of eq.~(\ref{s}) (see \citealt{b6} and \citealt{b2} for a detailed discussion).

Although the result (\ref{s}) was obtained for the $2 \times 2$ propagator of the matter density-velocity system, it holds, taking into account proper modifications, also when halos are included, {\it i.e.} for the $3\times 3$ propagator considered in this Section. As in \citet{b6}, we obtain an improved propagator interpolating between the 1-loop result (eq.~\ref{r}) at low $k$ and the Gaussian decay
(\ref{s}) (with a modified pre-factor) at high $k$. The details of the derivation and the relevant formulae are given in appendix B.

\subsection{TRG}
Unlike the propagator, the power spectrum cannot be resummed analytically at large k. 
Different semi-analytical procedures (\citealt{b5}, \citealt{b39}, \citealt{b43}) have been proposed to compute it in the mildly non-linear regime. In this paper we will consider the TRG technique introduced in \citet{b1}. 

Starting from eq. (\ref{e}), a hierarchy of differential equations for the power
spectrum, the bispectrum and higher order correlations is obtained. We choose
to truncate it at the level of the trispectrum $Q_{abcd}=0$, so that the
equations for $P_{ab}$ and $B_{abc}$ form a closed system
\begin{eqnarray}
&&\partial_\eta\,P_{ab}(k; \eta) =\nonumber\\
&&\quad -\, \Omega_{ac} (\eta)P_{cb}(k; \eta)  - \Omega_{bc} (\eta)P_{ac}(k; \eta) \nonumber\\
&&\quad + e^\eta \int d^3 q\, \left[ \gamma_{acd}({\bf k},\,{\bf -q},\,{\bf q-k})\,B_{bcd}({\bf k},\,{\bf -q},\,{\bf q-k};\,\eta)\right.\nonumber\\
&&\quad\left. + B_{acd}({\bf k},\,{\bf -q},\,{\bf q-k};\,\eta)\,\gamma_{bcd}({\bf k},\,{\bf -q},\,{\bf q-k})\right]\,,\nonumber\\
&&\nonumber\\
&&\partial_\eta\,B_{abc}({\bf k},\,{\bf -q},\,{\bf q-k};\,\eta) = \nonumber\\
&&\qquad\quad - \Omega_{ad} (\eta)B_{dbc}({\bf k},\,{\bf -q},\,{\bf q-k};\,\eta)\nonumber\\
&&\qquad\quad- \Omega_{bd} (\eta)B_{adc}({\bf k},\,{\bf -q},\,{\bf q-k};\,\eta)\nonumber\\
&&\qquad\quad - \Omega_{cd} (\eta)B_{abd}({\bf k},\,{\bf -q},\,{\bf q-k};\,\eta)\nonumber\\
&&\qquad\quad + 2 e^\eta \left[ \gamma_{ade}({\bf k},\,{\bf -q},\,{\bf q-k}) P_{db}(q;\,\eta)P_{ec}(|{\bf k}-{\bf q}|;\,\eta)\right.\nonumber\\
&&\qquad\quad +\gamma_{bde}({\bf -q},\,{\bf q-k},\,{\bf k}) P_{dc}(|{\bf k}-{\bf q}|;\,\eta)P_{ea}(k;\,\eta)\nonumber\\
&&\qquad\quad +\left. \gamma_{cde}({\bf q-k},\,{\bf k},\,{\bf -q}) P_{da}(k;\,\eta)P_{eb}(q;\,\eta)\right]\,,
\label{m}
\end{eqnarray}
which integrated
gives the power spectra at any redshift and for any momentum scale.
The system (\ref{m}) consists of coupled differential equations which are
solved numerically, starting from given initial conditions, i.e. 
$P_{ab}(k;\,\eta_i)$ and $B_{abc}({\bf k},\,{\bf -q},\,{\bf
  q-k};\,\eta_i)$.\\ From eq. (\ref{l}), we can observe that $\Omega_{13}$ and
$\Omega_{23}$ are zero, which means that the evolution of $\delta_m$ and
$\theta$ is not modified, with respect to the original TRG formulation, by the
presence of $\delta_h$, as it is expected.

\section{Results}
\label{num}
\subsection{Propagator}
\label{propres}
\begin{figure}
\includegraphics[width = 3.5in,keepaspectratio=true]{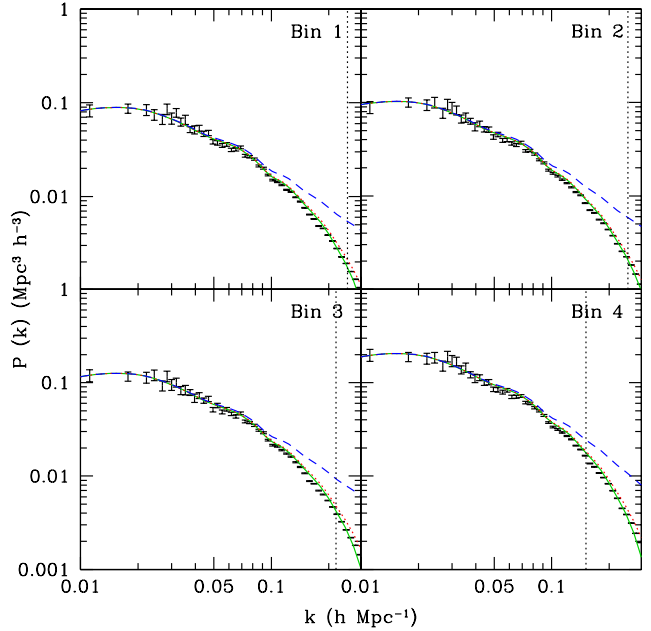}
\caption{Cross spectrum between the halo density at $z=0$ and the matter density at $z=50$. The outcome of the N-body simulation (black
  points with error bars) is compared against linear-theory $P_{PL}$ (blue dashed line) and resummed
  result $P_{PR}$ (green solid line). The red dotted line shows the effect of neglecting the non-Gaussian term, i.e. $P_{PR}-\Delta P^{NG}_{31}$.} The vertical black dotted lines represent the limit up to which we expect our fluid approximation for the haloes to work.
\label{fig9} 
\end{figure}

In order to assess the validity of our analytical approach, we compare our
results for the resummed propagator against the simulation; to this end, we
consider the relation in eq.~(\ref{ct}).
In particular we choose the indices $a~=~3,\, b=1$, so
that we can check the components related to the haloes that were not present in the 
original formalism by \citet{b6} and, at the same time, 
we do not have to deal with the shot-noise
problem. We extract the cross spectra from the simulation and compare them against those obtained both using linear theory propagators 
$P_{PL}\equiv g_{31} P^0_{11}+g_{32} P^0_{21}+g_{31} P^0_{31}$  and
resummed propagators $P_{PR}\equiv G_{31} P^0_{11}+G_{32} P^0_{21}+G_{31} P^0_{31}+ \Delta P^{NG}_{31}$; 
the result is shown in Figure \ref{fig9}.
We note that the linear model severely overpredicts the two-time cross spectrum for $k>0.05\, h\, \mathrm{Mpc}^{-1}$.
It is evident that the resummed theory improves considerably upon the linear
one, and agrees with the simulation within 10 per cent accuracy up to the scale where
the fluid approximation holds. We include in  $\Delta P^{NG}_{31}$ the effect of the initial non-Gaussianity of the halo field via its initial bispectra, computed as in eq.~(\ref{y}). It turns out that the components giving a non-vanishing effect are of the $B_{113}$ type. Their contribution is suppressed by a $D_+(z_{in})/D_+(z=0)$ factor with respect to that of an hypothetical primordial non-Gaussianity in the matter field (which we do not consider here). Therefore, the effect is of modest entity but, nevertheless, it improves the agreement with the simulation with respect to the case in which it is neglected.

\subsection{Power spectrum}
\begin{figure}
\includegraphics[width = 3.5in,keepaspectratio=true]{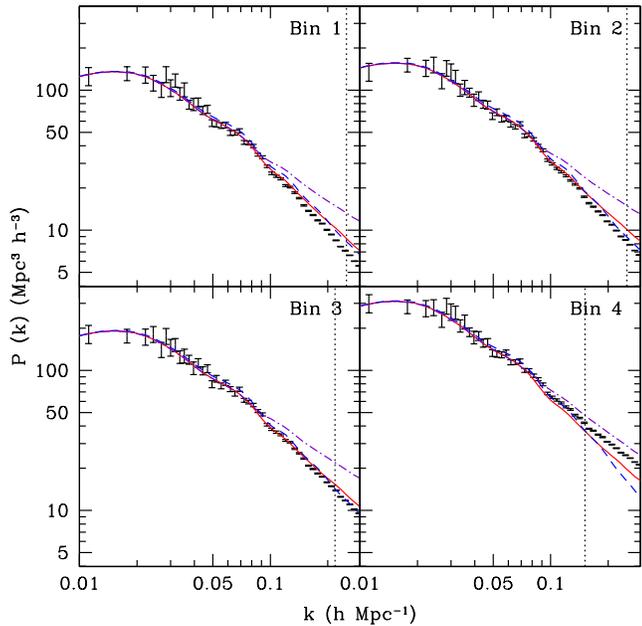}
\caption{The cross spectrum between matter and halo distribution at $z=0$ is shown in the four bins; the black dots with error bars represent the simulation, the blue dashed line is linear theory, the violet dot-dashed line is 1-loop and the red solid line is TRG. The vertical black dotted lines represent the limit up to which we expect our fluid approximation for the haloes to work.}
\label{fig3}
\end{figure}

\begin{figure}
\includegraphics[width = 3.5in,keepaspectratio=true]{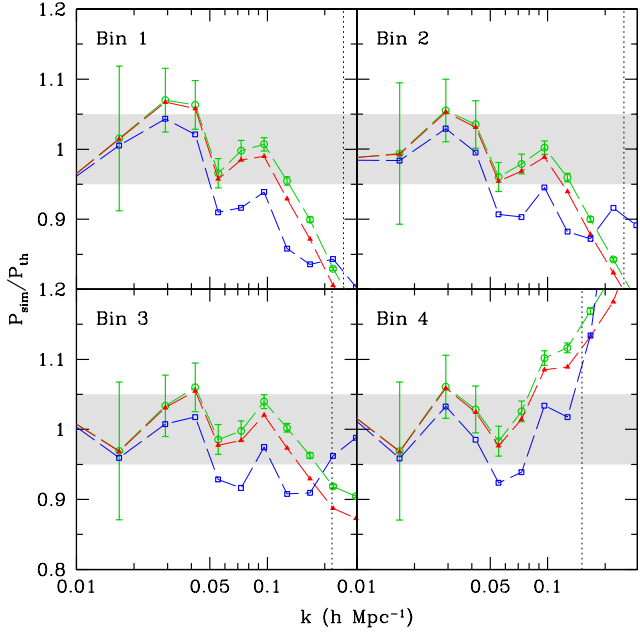}
\caption{The ratio between the spectrum from simulation and: linear theory (blue squares), TRG without bispectra (red triangles), TRG with bispectra (green circles). 
The vertical black dotted lines represent the limit up to which we expect our fluid approximation for haloes to work. The shaded area marks the 5 per cent accuracy interval.}
\label{fig5}
\end{figure}

\begin{figure}
\includegraphics[width = 3.5in,keepaspectratio=true]{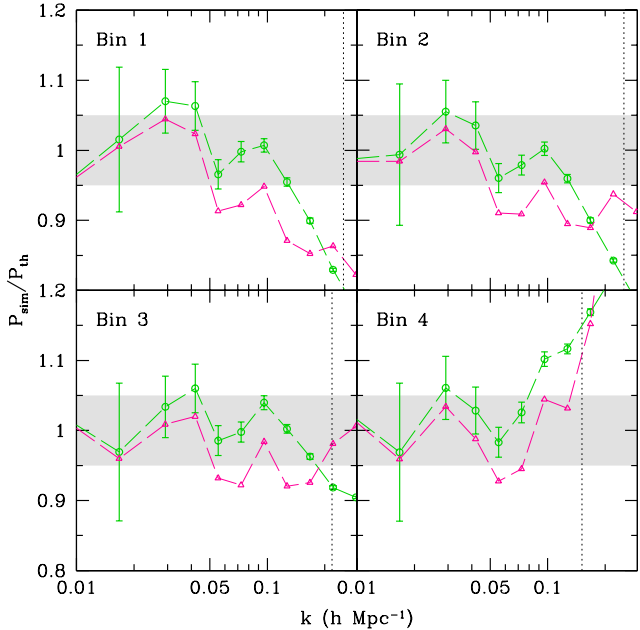}
\caption{The ratio between the cross spectrum from simulation and TRG with bispectra (green) and linear theory with the inclusion of velocity bias (magenta).
The vertical black dotted lines represent the limit up to which we expect our fluid approximation for haloes to work.}
\label{fig4}
\end{figure}

\begin{figure}
\includegraphics[width = 3.5in,keepaspectratio=true]{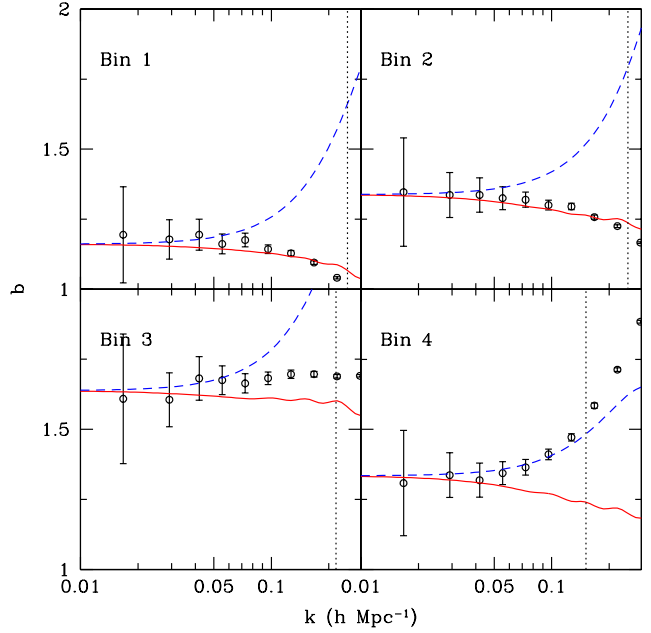}
\caption{The effective bias $b\equiv P_{mh}/P_m$ as a function of the wavevector in the four bins: the black circles with error bars are from the simulation, the blue dashed line represents linear theory and the red solid line TRG. The vertical black dotted lines represent the limit up to which we expect our fluid approximation for the haloes to work.}
\label{fig6}
\end{figure}

The TRG equations presented above are integrated numerically starting from the initial conditions discussed
in section \ref{init}. As a first step, we set all the initial
bispectra to zero
In Figure \ref{fig3} we show a comparison between the halo-matter cross spectra extracted from the N-body
simulation and the results of TRG, one-loop and linear theory. In the first three bins, corresponding to lower halo masses, linear theory overpredicts the power on mildly 
non-linear scales; note that this departure arises on larger scales compared to the matter auto spectrum (not shown in the figure). 
The overprediction of linear theory is cured by the 
one-loop power spectrum only on very large scale, while the TRG manages to correct it up to a smaller scale, before starting to fail. The fourth bin, though, displays a different behaviour: linear theory lacks of power on small scales, and neither the 1-loop correction nor the TRG are much more accurate. This might originate from the fact that very massive haloes are large and rare in the initial conditions, therefore less suited for the fluid approximation. Moreover, they also display the strongest velocity bias, which we are neglecting in our current non-linear treatment.
A more quantitative analysis is presented in Figure \ref{fig5}, where we plot the ratios between the spectra from the simulation and the theoretical results.
The TRG gives a cross spectrum within $5\%$ accuracy at least up to $k=0.1\, h\, \mathrm{Mpc}^{-1}$ (barring bin 4), while linear theory does so up to $k=0.05\, h\, \mathrm{Mpc}^{-1}$.\\
As previously pointed out, however, the halo-density field is not Gaussian at $z_{in}=50$. 
We use expressions (\ref{y}) to account for initial bispectra in the TRG method (they are not present in the linear theory and at 1-loop level). Figure \ref{fig5} illustrates that this indeed produces a slight improvement in the agreement with the simulation. 
The correction resulting from the introduction of the initial bispectra turns out to be quite small. The reason is that their contribution to the final cross spectrum $P_{13}$ is suppressed. To understand why, we can use perturbation theory; first, let us investigate the case of $P_{11}$. In the one-loop computation, the initial $P_{11}$ contributes to the final one with a ``weight" of $(D_+(z=0)/D_+(z_{in}))^2=e^{2\eta}$, as it also happens in linear theory. The initial matter bispectrum $B_{111}$, instead, has a weight of $e^{\eta}$, so it is suppressed by a factor of $e^{-\eta}$ \footnote{For the most experienced readers, this happens because one of the two vertices carrying the $e^{\eta}$ factor is replaced by the bispectrum.}. If we now consider the haloes, we have showed in (\ref{n}) that there is a new decaying mode, responsible for the linear debiasing. This new mode, that involves only the halo field, carries an extra $e^{-\eta}$ suppression factor. We can now rank the contributions to $P_{13}$ according to their relevance:
\begin{enumerate}
\item $P_{11}$
\item $P_{13}$, $ B_{111} $
\item $P_{33}$, $ B_{113} $
\item $B_{133}$
\item $B_{333}$.
\end{enumerate}
Each item is suppressed by a factor of $e^{-\eta}$ with respect to the previous one.
We can see that only $B_{111}$ has some relevance, while the other terms are highly suppressed. Even if the reasoning was based on perturbation theory, it is valid also for TRG, at a qualitative level. Incidentally, the fact that $P_{11}$ is the most relevant contribution is another evidence of debiasing.\\
We can now address the effect of the truncation we perform in the TRG, namely considering the trispectrum $Q=0$. First of all, the matter trispectrum $Q_{1111}$ can be neglected in the range of scales under consideration, as one can conclude from the comparison between TRG and simulations in \citet{b1}. The contribution from initial mixed (i.e. matter-halo) or pure halo trispectra is furtherly suppressed with respect to $Q_{1111}$ by extra $e^{-\eta}$ factors, for the same reason as above.
However, the trispectrum has its own time evolution as well, and one might argue that it becomes more relevant for $z<z_{in}$; this seems to be not the case, because enlarging the TRG truncation scheme by including the running of the trispectrum gives a contribution to the power spectrum which is at least of two-loop order and it is certainly subdominant in the scales we are considering.\\
From Section 5 onward, we neglected any velocity bias, since this approximation proved to be accurate enough at $z=0$ (in linear theory). As a further check, it is interesting to observe from Figure \ref{fig4} that the TRG is able to give a better prediction than full linear theory in eq. (\ref{vb}), even neglecting the velocity bias that the linear theory accounts for.\\
We can now look at the model predictions for the halo bias, defined as the ratio $P_{13}/P_{11}$. This quantity is plotted as a function of wavenumber in Figure \ref{fig6}.
While the linear-theory bias always increases with scale, irrespectively of halo mass, the TRG result closely follows the scale dependence of $b(k)$ seen in the simulation for the first two bins. It also gives a nearly constant bias for the third bin, as the simulation does, even though with a slightly lower value.
However, the linear model performs better in the last bin where the bias in the simulation increases with $k$.

\begin{figure}
\includegraphics[width = 3.5in,keepaspectratio=true]{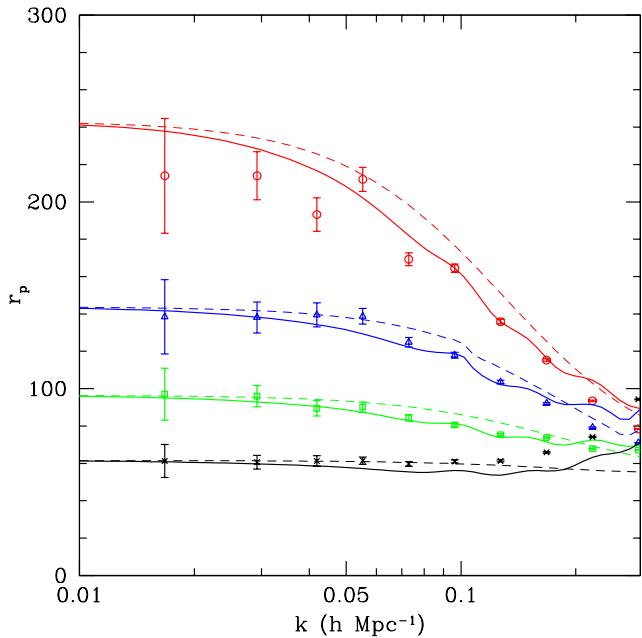}
\caption{The ratio between the halo-matter cross spectrum at $z=0$ and $z=50$ for the four mass bins: symbols represent the simulation, dashed lines the linear theory and solid lines TRG. From top to bottom, we have Bin 1 in red, Bin2 in blue, Bin 3 in green and Bin 4 in black. }
\label{fig7}
\end{figure}

\begin{figure}
\includegraphics[width = 3.5in,keepaspectratio=true]{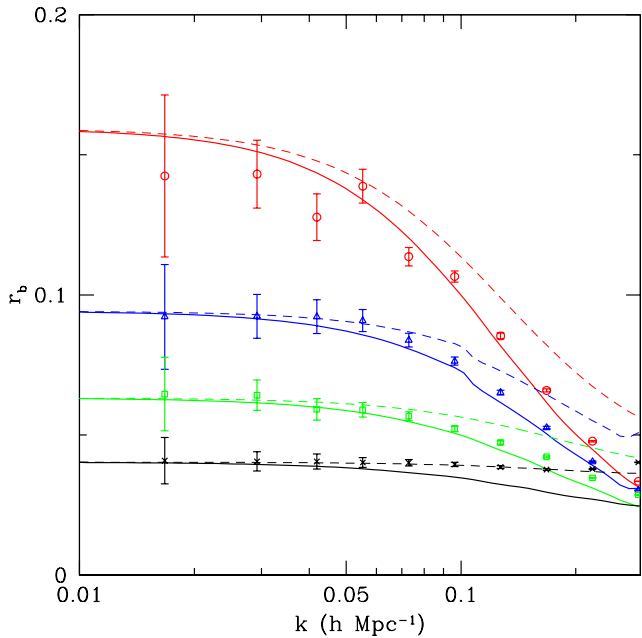}
\caption{The ratio between the bias at $z=0$ and $z=50$ for the four mass bins: symbols represent the simulation, dashed lines the linear theory and solid lines TRG. From top to bottom, we have Bin 1 in red, Bin2 in blue, Bin 3 in green and Bin 4 in black. }
\label{fig8}
\end{figure}

It is also interesting to look at the results displayed in a different way; we can investigate the evolution of the cross spectrum from the initial conditions to today and the evolution of the bias as well. In Figures \ref{fig7} and \ref{fig8} we plot, respectively, 
\begin{equation}
r_p\equiv \frac{P_{mh}(z=0)}{P_{mh}(z=50)} \qquad \mathrm{and} \qquad r_b\equiv \frac{b(z=0)}{b(z=50)}\,.
\end{equation}
Again, our model is able to match accurately the trend of the simulation and to improve upon linear theory, excluding bin 4. 
A key feature of the linear solution in eq. (\ref{uff}) is the debiasing between halo and matter distributions with time. 
It is worth noting from Figure \ref{fig8} that this effect is stronger for high-mass haloes, but constant on all the scales, while for low-mass haloes it is weaker on large scales and presents a strong $k$-dependence.

\section{Conclusions}
\label{con}
We have presented a novel approach to modeling the clustering of dark-matter haloes on mildly non-linear scales.
This follows the motion of the regions that will collapse to form haloes (that we dub proto-haloes).
Since the number of proto-haloes is conserved over time, for sufficiently large scales ($k<0.2\,h\,\mathrm{Mpc}^{-1}$), we can write a set of fluid equations that govern their evolution under the effect of gravity, which couples perturbations in the halo and matter density fields.
We provide analytical solutions for the linearized equations and 1-loop perturbative corrections for the halo and
matter power spectra. For the propagator, quantifying the memory of the density and velocity fields to their initial conditions, 
we also perform a resummation of perturbative corrections. Finally, for the power spectrum we compute the non-linear evolution 
using a semi-analytical procedure, namely an extension of the time renormalization group.

The initial conditions for our evolutionary equations are specified adopting a Lagrangian bias model, originally developed to describe the clustering and motion of linear density peaks.
We fix the parameters of the model so that to reproduce the distribution of proto-haloes in a high-resolution N-body simulation at $z=50$.
We use the same simulation to test the predictions of our model at $z=0$.

Our main results can be summarized as follows:
\begin{itemize}
\item Independently of the initial conditions, in the linear solution the halo density and velocity fields asymptotically match the corresponding matter fields at late times.
This 'debiasing' develops at a different rate for the density and the velocity, being faster for the latter.
\item Even if there is no initial density bias, the presence of a velocity bias generates a transient density bias that vanishes at late times.
\item Neglecting any initial velocity bias alters the linear predictions for the halo-matter cross spectrum at redshift $z=0$ only by less than 3 per cent, for $k<0.3\,h\,\mathrm{Mpc}^{-1}$. This provides us with the motivation to ignore the velocity bias in the non-linear analysis.
\item Unlike its linear counterpart, the resummed propagator is in good agreement with the N-body simulation, independently of halo mass.
\item The halo-matter cross spectrum predicted by the TRG is accurate to $5$ per cent up to $k\simeq 0.1\, h\,\mathrm{Mpc}^{-1}$ for a broad range of halo masses.
This does not hold for very massive haloes ($M>10^{14} h^{-1} M_{\odot}$), that have low number density and high initial velocity bias, for which discreteness effects are more important.
\item The TRG result improves upon both linear theory and 1-loop corrections. Its performance is slightly enhanced accounting for the initial non-Gaussianity of the halo distribution.
\item For low halo masses our model accurately describes the scale-dependent bias measured in the simulation at $z=0$.
\end{itemize}

\section*{Acknowledgments}
AE was supported through the SFB-Transregio 33 ``The Dark Universe" by  
the Deutsche Forschungsgemeinschaft (DFG).\\
AE and SK acknowledge BCGS support for part of the work.\\
CP thanks the participants to the workshop ``Modern Cosmology: Early Universe, CMB and LSS" held at the Centro de Ciencias de Benasque Pedro Pascual 
for stimulating discussions. Part of this work has been developed there in August 2010.

\appendix

\onecolumn
\section[]{Complete analytic expressions of the 1-loop propagator and power
  spectrum}
\noindent{Linear power spectra:}
\begin{eqnarray}
P_{L,11}(k;\eta,\eta)&=&P^0_{11}(k)=P_{L,12}(k;\eta,\eta)=P_{L,22}(k;\eta,\eta)\,,\nonumber\\
P_{L,13}(k;\eta,\eta)&=&e^{-\eta}((e^{\eta}-1)P^0_{11}(k)+P^0_{13}(k))\,,\nonumber\\
P_{L,33}(k;\eta,\eta)&=&e^{-2\eta}((e^{\eta}-1)^2P^0_{11}(k)+P^0_{33}(k)+2(e^{\eta}-1)P^0_{13}(k))\,.
\end{eqnarray}

\subsection{Propagator}

First we report the integrands of the one-loop corrections of the components
we are interested in.
The linear power spectrum can be split as
\beq
P_{L,c_i,c_j}(q_i;s_i,s_j) = P^0(q_i) U_{c_i,c_j} + \Delta
P_{c_i,c_j}(q_i;s_i,s_j)\,,
\label{app1}
\eeq
where 
\beq
U_{c_i,c_j} = 1 \quad  \mathrm{for\; any} \;\; c_i,\,c_j\,,
\label{app2}
\eeq
while
\beq
\Delta P_{c_i,c_j}(q_i;s_i,s_j) \neq 0  \quad  \mathrm{only\; if} \;\; c_i \;\mathrm{or}\;c_j=3 \,.
\label{bw}
\eeq
Looking at eq. (\ref{r}) we can consistently split the one-loop correction to the propagator as
$\Delta g_{ab}=\Delta^{(1)} g_{ab}+\Delta^{(2)} g_{ab}$ and we denote them as $\delta^{(i)} g_{ab}$, so that
$\Delta^{(i)}  g_{ab}=\int \mathrm{d}q\, \delta^{(i)}  g_{ab}$, for $i=1,2$. 
\begin{eqnarray}
\delta g_{11}&=&\frac{\pi \left(e^{\eta }-1\right)^2 P^0_{11}(q) \left[4 \left(6 k^7
   q-79 k^5 q^3+50 k^3 q^5-21 k q^7\right)-3
   \left(k^2-q^2\right)^3 \left(2 k^2+7 q^2\right) \log
   \left(\frac{(k+q)^2}{(k-q)^2}\right)\right]}{840 k^3 q^3}\,,\nonumber\\
\delta^{(1)} g_{31}&=&\delta g_{11}\,,\nonumber\\
\delta^{(1)} g_{32}&=&\frac{2}{3} \delta g_{11}\,,\nonumber\\
\delta^{(1)} g_{33}&=&-\frac{2}{3} e^{-\eta } (e^{\eta} -1)^2 k^2 \pi P^0_{11}(q)\,,\nonumber\\
\delta^{(2)} g_{31}&=&\frac{\pi  e^{\eta } \left(P^0_{13}(q)-P^0_{11}(q)\right) \left(4 \left(9 k^5 q+24 k^3 q^3-9 k q^5\right)-18 \left(k^2-q^2\right)^3 \log
   \left(\left|\frac{k+q}{k-q}\right|\right)\right)}{280 k^3 q}\,,\nonumber\\
\delta^{(2)} g_{32}&=&\frac{2}{3} \delta^{(2)} g_{31}\,,\nonumber\\
\delta^{(2)} g_{33}&=& 0
\end{eqnarray}

\subsection{1-loop PS}
From eq. (\ref{phi}), one can see that the one-loop corrections require integrals
over $\mathrm{d}s_1$, $\mathrm{d}s_2$ and $\mathrm{d}^3q= q^2 \sin\theta \,\mathrm{d}q\,
\mathrm{d}\theta\, \mathrm{d}\phi$; the integration over $q$ and over $\mu\equiv\cos\theta$ cannot be
performed analytically. The following expressions are therefore integral
kernels, denoted by $\delta_{q\mu}P$ (the factor of $q^2$ is already taken
into account).
If $\cos\theta=\mu=\frac{{\bf k}\cdot{\bf q}}{k\,q}$ and $|k-q|=\sqrt{k^2-2 k \mu  q+q^2}$,
\begin{eqnarray}
\delta_{q\mu} P^{II\phantom{ab}}_{11}&=&\frac{\pi  k^4 e^{2 \eta } P^0_{11}(q) \left[7 k \mu +\left(3-10 \mu ^2\right) q\right]^2 P^0_{11}\left(|k-q|\right)}{49 \left(k^2-2 k \mu  q+q^2\right)^2}\,,\nonumber\\
\end{eqnarray}
\beqra
\delta_{q\mu} P^{II\phantom{ab}}_{13}&=&\frac{\pi  k^3 e^{\eta } \left(7 k \mu
    +\left(3-10 \mu ^2\right) q\right)}{49 \left(k^2-2 k \mu q+q^2\right)^2}
\left[7 q (k-\mu  q) P^0_{13}(q) P^0_{11}\left(|k-q|\right)+P^0_{11}(q) 7 \mu  \left(k^2-2 k \mu  q+q^2\right)
    P^0_{13}\left(|k-q|\right)\right.\nonumber\\
&&\left.+ k P^0_{11}(q) P^0_{11}\left(|k-q|\right)
   \left(7 k \mu  \left(e^{\eta }-1\right)-q \left(\left(10 \mu ^2-3\right)
       e^{\eta }-14 \mu ^2+7\right)\right)\right]\,.
\eeqra
\begin{eqnarray}
\delta_{q\mu} P^{II\phantom{ab}}_{33}&=&\frac{\pi  k^2}{49 \left(k^2-2 k \mu  q+q^2\right)^2} \left\{98 \mu  q (k-\mu  q) \left(k^2-2 k \mu
    q+q^2\right) P^0_{13}(q) P^0_{13}\left(|k-q|\right)\right.\nonumber\\
&&+ 7 q (k-\mu  q) P^0_{11}\left(|k-q|\right) \left[2 k P^0_{13}(q) \left[7 k \mu  \left(e^{\eta }-1\right)-q \left(\left(10 \mu ^2-3\right) e^{\eta }-14 \mu ^2+7\right)\right]+7 q P^0_{33}(q) (k-\mu 
   q)\right]\nonumber\\
&&+k^2 P^0_{11}(q) P^0_{11}\left(|k-q|\right) \left[q \left(\left(10 \mu ^2-3\right) e^{\eta
    }-14 \mu ^2+7\right)-7 k \mu  \left(e^{\eta }-1\right)\right]^2\nonumber\\
&&+49 \mu ^2 P^0_{11}(q) \left(k^2-2 k \mu  q+q^2\right)^2
P^0_{33}\left(|k-q|\right)\nonumber\\
&&\left.+14 k \mu  P^0_{11}(q) \left(k^2-2 k \mu  q+q^2\right) \left[7 k \mu  \left(e^{\eta }-1\right)-q \left(\left(10 \mu ^2-3\right) e^{\eta }-14 \mu ^2+7\right)\right] P^0_{13}\left(|k-q|\right)\right\}\,.
\end{eqnarray}

\section[]{The resummed propagator}
As it was discussed in \citet{b6}, the leading contribution in $k^2 \exp(2\eta)$ to the propagator at a generic $n$-loop order contains a chain of propagators and vertices of the form
\beq
g_{a\,b_1}(\eta-s_1) \gamma_{b_1\,c_1\,a_1}(\bk,-\bq_1,\bq_1-\bk) g_{a_1\,b_2}(s_1-s_2)\cdots  
g_{a_{2n-1}\,b_{2n}}(s_{2n-1}-s_{2n}) \gamma_{b_{2n}\,c_{2n}\,a_{2n}}(\bk+\bq_{2n},-\bq_{2n},-\bk) g_{a_{2n}\,b}(s_{2n})\,.
\label{chain}
\eeq
The $c_i$ indices have to be contracted in all possible pairings, by inserting $n$ linear power spectra, 
each of the form
\beq
\delta_D(\bq_i+\bq_j) P_{L,c_i,c_j}(q_i;s_i,s_j)\,.
\eeq
The $n$-loop contribution is obtained by multiplying by $\exp(\sum_{i=1}^{2n} s_i)$ and by the appropriate combinatoric factor, and then by integrating over
\beq
\Pi_{i=1}^{2n} \left(\int_0^\eta ds_i \int d^3 q_i \right)\,.
\eeq
Recall eqs. (\ref{app1}-\ref{bw}): if all the insertions are of the $P^0(q_i) U_{c_i,c_j}$ type then the resummation goes exactly as in the standard case. Indeed
\beq
 U_{c_1,c_j} g_{a b_1}(\eta-s_1) \gamma_{b_1\,c_1\,a_1}(\bk,-\bq_1,\bq_1-\bk) g_{a_1\,b_2}(s_1-s_2) \to u_{c_j} g_{a\,b_2}(\eta-s_2) \frac{1}{2}\frac{\bq \cdot \bk}{q^2}\,,
\eeq
in the $k \gg q$ limit. Besides the explicit form for the vertices (see Section ~1), \
we have used the composition property of the propagators
\beq
 g_{a b}(\eta-s_1) g_{b c}(s_1-s_2)= g_{a c}(\eta-s_2)\,,
\eeq
and have defined $u= (1,1,1)$.
Since each vertex, contracted by a $u_{c_j}$ vector becomes proportional to a delta function in its first and third index, the chain of propagators composes up to a single propagator $g_{a b}(\eta)$ and the time integral can be easily performed. The momentum integrals factorize into $n$ integrals of the type
\beq
- \int d^3 q P^0(q)    \frac{(\bk\cdot \bq)^2}{q^4} = - k^2 \sigma^2\,,
\eeq
where $\sigma^2$ has been defined in (\ref{t}). 
Using the appropriate combinatoric factors, the leading contribution to the propagator at $n$-loop, in the large momentum limit, when all the power spectrum insertions are of the $ P^0(q_i) U_{c_i,c_j} $ type is
\beq
\frac{1}{n!} \left[-k^2 \sigma^2\frac{(e^\eta -1)^2}{2}\right]^n g_{ab}(\eta)\,,
\label{nord}
\eeq
which resums to $g_{ab}(\eta) \exp[-k^2 \sigma^2\frac{(e^\eta -1)^2}{2}]$.

As for the insertions including the $ \Delta P_{c_i,c_j}(q_i;s_i,s_j)$ contribution to the linear power spectrum, 
the important point to realize is that, due to (\ref{bw}) and to the structure 
of the linear propagator and the vertices, 
a chain like (\ref{chain})  can be contracted by {\em at most one} $ \Delta P_{c_i,c_j}(q_i;s_i,s_j)$ power spectrum,
the remaining ones being of the type $P^0(q_i) U_{c_i,c_j}$. 
Moreover, these insertions only contribute to $G_{31}$ and $G_{32}$. 
Therefore, at $n$-th order, we have only two types of contributions: those with all $P^0(q_i) U_{c_i,c_j}$ insertions, giving (\ref{nord}), and those with one  $ \Delta P_{c_i,c_j}$ insertions and $n-1$ $P^0(q_i) U_{c_i,c_j}$ ones, which can also be resummed in the large $k$ limit.

As a consequence, the complete resummed propagator in the large momentum limit is
\beq
G_{ab}(k;\eta) = (g_{ab}(\eta) + \delta_{a3} f_b \Delta^{(2)} g_{31}(k;\eta))\exp[-k^2 \sigma^2\frac{(e^\eta -1)^2}{2}]\,,
\eeq
with $f_b=(1,\,2/3,\,0)$.

In order to have an expression for the propagator valid at any k, one can proceed as in (\cite{b6}) and interpolate between the large $k$ limit above and the 1-loop result at low k. This can be done, for instance for $G_{11}$, starting from its 1-loop expression
\beq
g_{11}(\eta)+\Delta g_{11}(k;\eta) \simeq g_{11}(\eta)\left(1 +\frac{5}{3} \Delta g_{11}(k;\eta) \right)\,,
\eeq
where the above approximation is exact in the large $\eta$ limit. Since 
\beq
\frac{5}{3} \Delta g_{11}(k;\eta)  \to -k^2 \sigma^2 \frac{(e^\eta-1)^2}{2}\,,\qquad\mathrm{for\;large}\;k\,,
\eeq
the required interpolation is given by
\beq
G_{11}(k;\eta)=g_{11}(\eta)  \exp\left[\frac{5}{3} \Delta g_{11}(k;\eta) \right]\,.
\eeq
Proceeding analogously for the other components, we have:
\beqra
G_{12}(k;\eta)&=&g_{12}(\eta)  \exp\left[\frac{5}{3} \Delta g_{11}(k;\eta) \right]\,,\nonumber\\
G_{22}(k;\eta)&=&g_{22}(\eta)  \exp\left[\frac{5}{2} \Delta g_{22}(k;\eta) \right]\,,\nonumber\\
G_{21}(k;\eta)&=&g_{21}(\eta)  \exp\left[\frac{5}{2} \Delta g_{22}(k;\eta) \right]\,,\nonumber\\
G_{31}(k;\eta)&=&\left(g_{31}(\eta)+\Delta^{(2)} g_{31}(k;\eta) \right)  \exp\left[\frac{5}{3} \Delta g_{11}(k;\eta) \right]\,,\nonumber\\
G_{32}(k;\eta)&=&\left(g_{32}(\eta) +\frac{2}{3}\Delta^{(2)} g_{31}(k;\eta) \right)  \exp\left[\frac{5}{3} \Delta g_{11}(k;\eta) \right]\,,\nonumber\\
G_{33}(k;\eta)&=&g_{33}(\eta) \exp\left[e^\eta \Delta^{(1)} g_{33}(k;\eta) \right]\,,
\eeqra
where we have used the identities
\beqra
 \Delta g_{12}(k;\eta)& =&\frac{2}{3} \Delta g_{11}(k;\eta)\,,\nonumber\\
 \Delta g_{21}(k;\eta) &=&\frac{3}{2} \Delta g_{22}(k;\eta)\,,\nonumber\\
 \Delta^{(1)} g_{31}(k;\eta) &=&\Delta g_{11}(k;\eta)\,\nonumber\\
  \Delta^{(1)} g_{32}(k;\eta) &=&\frac{2}{3} \Delta^{(1)} g_{31}(k;\eta)\,\nonumber\\
\Delta^{(2)} g_{32}(k;\eta)&=&\frac{2}{3}\Delta^{(2)} g_{31}(k;\eta)\,.
\eeqra

\section[]{Taking into account the initial halo non-Gaussianity.}
\label{ang}
While at $z_{in}$ the matter field can be considered gaussian with high accuracy, the same not necessarily holds for haloes. An initial non-Gaussianity for the equal-time power spectrum can be easily incorporated in the TRG formalism as discussed in \citet{b44}, and has been done in sect. 6.2. In order to obtain the effect of an initial non-Gaussianity on the cross-correlator at different times considered in eq.~(\ref{ct}), it suffices to compute the $O(\gamma)$ correction to the linear evolution of the field, by inserting eq.~(\ref{new3}) in (\ref{e}), to get 
\beq
\vp_a^{(1)}(\bk;\eta) = \int_0^\eta ds \,g_{ac}(\eta-s)e^s \int d^3 q\, \gamma_{cde}(\bk,-\bq,\bq-\bk) g_{df}(s) \vp_f(\bq;0)  g_{eg}(s) \vp_g(\bk-\bq;0)\,.
\eeq
The cross-correlator 
\beq
\langle \vp_a^{(1)}(\bk;\eta) \vp_b (\bk^\prime;0)\rangle \,,
\eeq
then includes the non-Gaussian expectation value
\beq
\langle \vp_f(\bq;0) \vp_g(\bk-\bq;0) \vp_b (\bk^\prime;0)\rangle = \delta_D(\bk+\bk^\prime)  B_{fgb}(\bq,\bk-\bq,-\bk)\,,
\eeq
and therefore gives eq.~(\ref{eng}).

\bsp

\label{lastpage}

\end{document}